\documentclass[aps,prl,
   amsfonts,amssymb,amsmath,
   twocolumn,
   superscriptaddress,
   noshowpacs,
   preprintnumbers,
   amsmath,amssymb
]{revtex4-1}

\usepackage{color}
\usepackage{graphicx}
\usepackage[pdftex,colorlinks=true,linkcolor=red,citecolor=blue]{hyperref}

\begin{document}

\title{Electronic tuning and uniform superconductivity in CeCoIn$_{5}$}

\author{K. Gofryk}
\email{gofryk@lanl.gov}
\affiliation{Los Alamos National Laboratory, Los Alamos, New Mexico 87545, USA}
\author{F. Ronning}
\email{fronning@lanl.gov}
\affiliation{Los Alamos National Laboratory, Los Alamos, New Mexico 87545, USA}
\author{J.-X.~Zhu}
\affiliation{Los Alamos National Laboratory, Los Alamos, New Mexico 87545, USA}
\author{M.~N.~Ou}
\affiliation{Los Alamos National Laboratory, Los Alamos, New Mexico 87545, USA}
\affiliation{Institute of Physics, Academia Sinica, Taipei, Taiwan}
\author{P.~H.~Tobash}
\affiliation{Los Alamos National Laboratory, Los Alamos, New Mexico 87545, USA}
\author{S.~S.~Stoyko}
\affiliation{Department of Chemistry, University of Alberta, Edmonton, Alberta T6G 2G2 Canada}
\author{X.~Lu}
\affiliation{Los Alamos National Laboratory, Los Alamos, New Mexico 87545, USA}
\author{A.~Mar}
\affiliation{Department of Chemistry, University of Alberta, Edmonton, Alberta T6G 2G2 Canada}
\author{T.~Park}
\affiliation{Department of Physics, Sungkyunkwan University, Suwon 440-746, South Korea}
\author{E.~D.~Bauer}
\affiliation{Los Alamos National Laboratory, Los Alamos, New Mexico 87545, USA}
\author{J.~D.~Thompson}
\affiliation{Los Alamos National Laboratory, Los Alamos, New Mexico 87545, USA}
\author{Z.~Fisk}
\affiliation{Department of Physics and Astronomy, University of California, Irvine, California 92697, USA}

\date{\today}

\begin{abstract}

We report a globally reversible effect of electronic tuning on the magnetic phase diagram in CeCoIn$_{5}$ driven by electron (Pt and Sn) and hole (Cd, Hg) doping. Consequently, we are able to extract the superconducting pair breaking component for hole and electron dopants with pressure and co-doping studies, respectively. We find that these nominally non-magnetic dopants have a remarkably weak pair breaking effect for a $d$-wave superconductor. The pair breaking is weaker for hole dopants, which induce magnetic moments, than for electron dopants. Furthermore, both Pt and Sn doping have a similar effect on superconductivity despite being on different dopant sites, arguing against the notion that superconductivity lives predominantly in the CeIn$_{3}$ planes of these materials. In addition, we shed qualitative understanding on the doping dependence with density functional theory calculations.

\end{abstract}

\maketitle

In a superconductor whose order parameter changes sign, non-magnetic impurities are expected to be strongly pair breaking and will suppress $T_{c}$ rapidly, similar to the effect of magnetic impurities in conventional superconductors\cite{bal}. Indeed, a few percent of Zn impurities in YBCO causes a dramatic suppression of $T_{c}$\cite{all}. However, the lack of $T_{c}$ suppression by various dopants in Fe-based superconductors has been argued as evidence that the order parameter does not change sign \cite{Kontani}. The heavy fermion superconductors Ce$T$In$_{5}$ ($T$ = Co, Rh, Ir) are prototypical of a general class of strongly correlated systems in which unconventional superconductivity (SC) emerges in close proximity to an antiferromagnetic quantum critical point (QCP) as in the cuprates and pnictides. The pure compound CeCoIn$_{5}$ is a high temperature (relative to $E_{F}$) $d$-wave superconductor\cite{2}, and consequently provides a great opportunity to study the effects of impurities on an unconventional superconductor.

\begin{figure}[t!]
\begin{centering}
\includegraphics[width=0.45\textwidth]{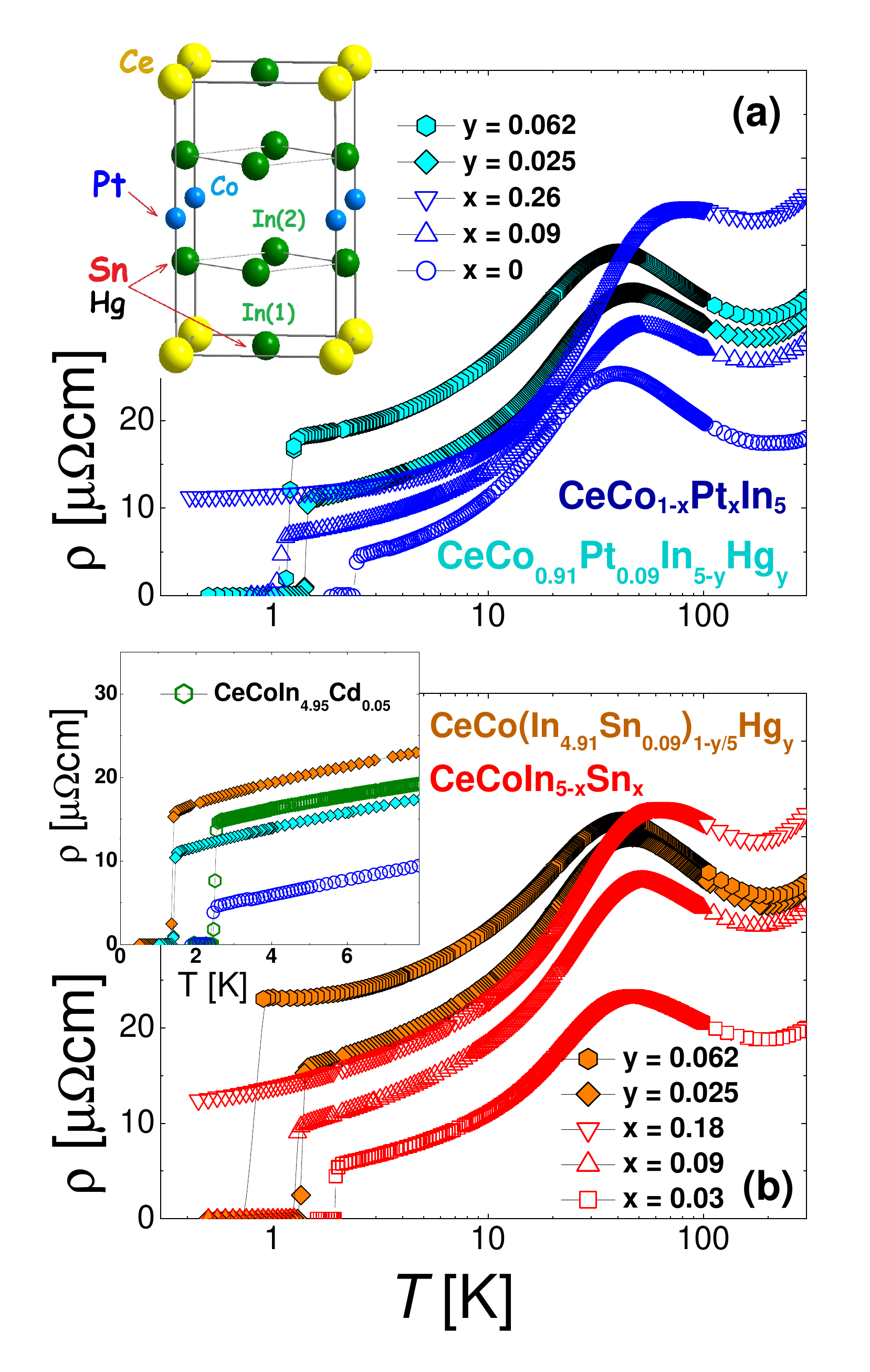}
\caption{(Color online) (a,b): The temperature dependence of the electrical resistivity of CeCoIn$_{5-x}$Sn$_{x}$, CeCo$_{1-x}$Pt$_{x}$In$_{5}$, CeCo$_{0.91}$Pt$_{0.09}$In$_{5-y}$Hg$_{y}$ and CeCo(In$_{4.91}$Sn$_{0.09}$)$_{1-\frac{y}{5}}$Hg$_{y}$. Inset to (a) crystal structure of CeCoIn$_{5}$ which can  be viewed as an alternating series of 'active' CeIn$_{3}$ and 'buffer' $T$In$_{2}$ layers; and (b) low temperature resistivity of CeCoIn$_{5}$ and the optimally tuned co-doped samples. Also shown is CeCoIn$_{4.95}$Cd$_{0.05}$ at $p$ = 1.6 GPa.}\label{fig.1}
\end{centering}
\end{figure}

Due to the small energy scales, however, heavy fermions are also easily tuned electronically. Indeed, hole doping with only 1\% of Cd or Hg induces an antiferromagnetic ground state in CeCoIn$_{5}$ \cite{11,14}, while electron doping with Sn has the opposite effect to hole doping, suppressing $T_{c}$, and eventually recovering a Fermi liquid ground state \cite{15,16,17,18,19}. It has been argued that the reason for the strong doping dependence is due to the fact that the dopant atoms Sn, Cd, and Hg preferentially substitute on the In(1) site in the 'active' CeIn$_{3}$ planes\cite{16} (see structure in Fig.~\ref{fig.1}a). Indeed, in CeCo$_{1-x}$Rh$_{x}$In$_{5}$, superconductivity persists up to 75\% Rh\cite{20}. However, the only substitutions that have been made in the $T$In$_{2}$ 'buffer' layers have been isoelectronic.

In this Letter, we show that Pt doping on the transition metal site is similar to Sn doping, both of which contribute 1 electron per dopant atom. Thus, the strength of pair breaking created by electron doping is roughly independent of the dopant location within the unit cell. By pressure\cite{11} and by co-doping with Hg [this work] we find that the magnetic phase diagram and the coherence temperature $T^{*}$ are reversible via local electronic tuning, enabling a determination of the pair breaking scattering rates.  While the pair breaking for electron dopants is stronger than for hole dopants, both are strikingly much weaker than expected for nominally non-magnetic impurities in a $d$-wave superconductor \cite{AG}. In addition, we show that density functional theory calculations qualitatively describe the observed doping dependence in this Ce-based heavy fermion system.

All single crystals were synthesized using an indium self-flux method\cite{2,21,17}. The doping levels were determined by single crystal x-ray diffraction and by microprobe analysis. For the physical properties measurements of CeCoIn$_{5-x}$Sn$_{x}$, crystals from the same batch as in Ref.~\onlinecite{17} have been used. For Pt and Hg doping the actual concentrations have been established to be 85~\% and 16~\% of the nominal concentrations, respectively. The actual concentrations are used in this Letter. The electrical resistivity were measured using a four wire ($AC$) method, implemented in a Quantum Design PPMS-9 device.

\begin{figure}[t!]
\begin{centering}
\includegraphics[width=0.45\textwidth]{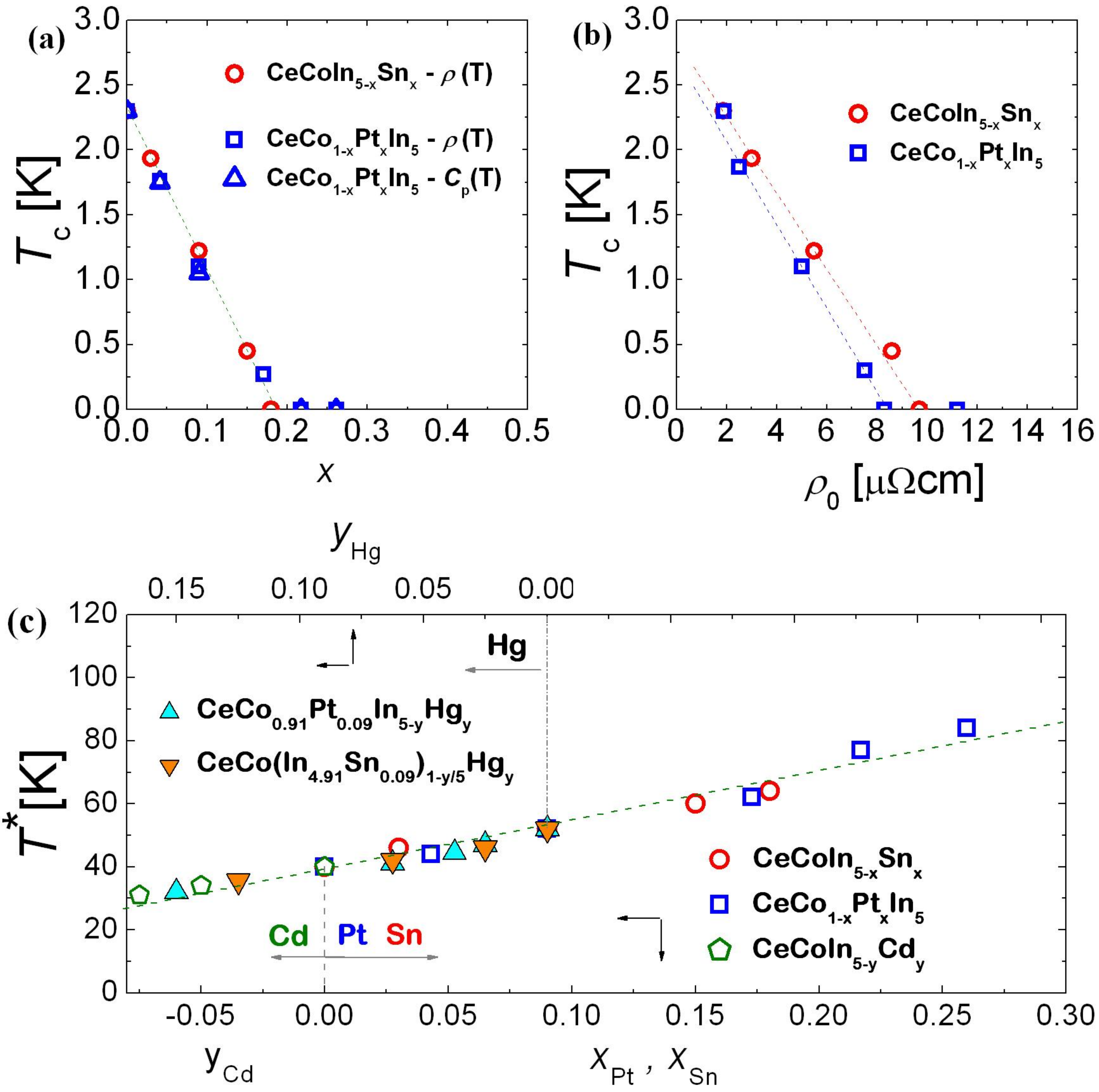}
\caption{(Color online) (a) Temperature-composition phase diagram and (b) $T_{c}$ vs. $\rho_{0}$ for CeCoIn$_{5-x}$Sn$_{x}$ and CeCo$_{1-x}$Pt$_{x}$In$_{5}$. (c) $T^{*}$~vs.~$x$ and $y$ for CeCoIn$_{5-x}$Sn$_{x}$, CeCoIn$_{5-y}$Cd$_{y}$ CeCo$_{1-x}$Pt$_{x}$In$_{5}$, CeCo$_{0.91}$Pt$_{0.09}$In$_{5-y}$Hg$_{y}$ and CeCo(In$_{4.91}$Sn$_{0.09}$)$_{1-\frac{y}{5}}$Hg$_{y}$.}\label{fig.2}
\end{centering}
\end{figure}

The main panels in Fig.~\ref{fig.1}a and \ref{fig.1}b show the temperature dependencies of the electrical resistivity of selected samples of CeCo$_{1-x}$Pt$_{x}$In$_{5}$, CeCo$_{0.91}$Pt$_{0.09}$(In$_{1-y}$Hg$_{y}$)$_{5}$, CeCoIn$_{5-x}$Sn$_{x}$ and CeCo(In$_{4.91}$Sn$_{0.09}$)$_{1-\frac{y}{5}}$Hg$_{y}$. All the curves show behavior typical of heavy-fermion materials as characterized by a pronounced maximum at the coherence temperature $T^{*}$. Such behavior is related to the onset of coherent Kondo screening of the magnetic moments by the conduction electrons at $T^{*}$. For superconducting samples the superconductivity is clearly visible as marked by a sharp drop of $\rho$ to zero at $T_{c}$ which agree with the bulk transitions measured by specific heat (not shown). As can be seen from Figs~\ref{fig.1} and ~\ref{fig.2}a, the substitution with Pt leads to a $T_{c}$ suppression similar to the one observed in Sn doping\cite{15,16,17,18,19}. As seen from Fig.~\ref{fig.2}b, superconductivity is completely suppressed at about 10~$\mu\Omega cm$ which is about half that observed for rare-earth doped CeCoIn$_{5}$\cite{3}.

All dopants of CeCoIn$_{5}$ have a pair breaking and an electronic tuning component. The electronic tuning is evident in Fig.~\ref{fig.2}c where electron doping (both Pt and Sn) increases the coherence temperature $T^{*}$ while hole doping (Cd and Hg) has the reverse effect. The electronic tuning is also reflected in the low temperature magnetic phase diagram where electron doping drives the system away from quantum criticality towards a paramagnetic state while hole doping pushes the system into a long range antiferromagnetically ordered state \cite{11,14}.

Our density functional theory (DFT) calculations reveal that this electronic tuning effect arises from a local change in the hybridization strength of the Ce $f$-electrons to the conduction electrons caused by the dopant atoms. We use a supercell which is eight times the size of the conventional unit cell (see the supplementary information for details). Fig.~\ref{fig.3} shows the resulting partial density of states (PDOS) for CeCoIn$_{5}$ doped with 2.5\% of Cd, Hg, and Sn. The PDOS of In and Ce sites which lie far from the impurity atom are virtually identical to that of pure CeCoIn$_{5}$. Meanwhile, the Sn (Cd,Hg) $p$-states possess a larger (smaller) bandwidth and are shifted down (up) in energy relative to the In states they replaced. A larger bandwidth reflects a larger hybridization. When one compares the Ce $f$-orbital PDOS for undoped and doped cases, it can be seen that the PDOS for Ce close to the impurity (labeled Ce1) is slightly broader (narrower) with Sn (Cd) doping relative to the Ce sites further away from the dopant atom (labeled Ce2), suggestive of an enhanced hybridization in the Sn doped case. While these calculations do not capture the many body physics necessary to describe the low energy properties of Ce-based heavy fermions, they provide a qualitative understanding of the doping trends. In the Doniach picture as the Kondo coupling grows the coherence temperature is expected to rise and simultaneously magnetic order becomes suppressed\cite{Doniach}. The expression for the Kondo coupling is $J_K$ = $V^2/(\epsilon_f - E_F)$ where $V$ is the hybridization to the bare $f$-level with energy $\epsilon_f$ and $E_F$ is the Fermi energy \cite{SW}. The increased (decreased) hybridization seen for Sn (Cd, Hg) doping is consistent with the increased (decreased) coherence temperature and suppression (onset) of magnetism observed experimentally. It is also interesting to note that the calculations report a near identical bare electronic structure for Cd and Hg doped CeCoIn$_{5}$, consistent with the experimentally identical phase diagrams\cite{11,14}. To uncover the full microscopic physics underlying the electronic tuning effects, a more quantitative theoretical analysis of hybridization between Ce and ligand atoms is required, and left for future work.

\begin{figure}[t!]
\begin{centering}
\includegraphics[width=0.45\textwidth]{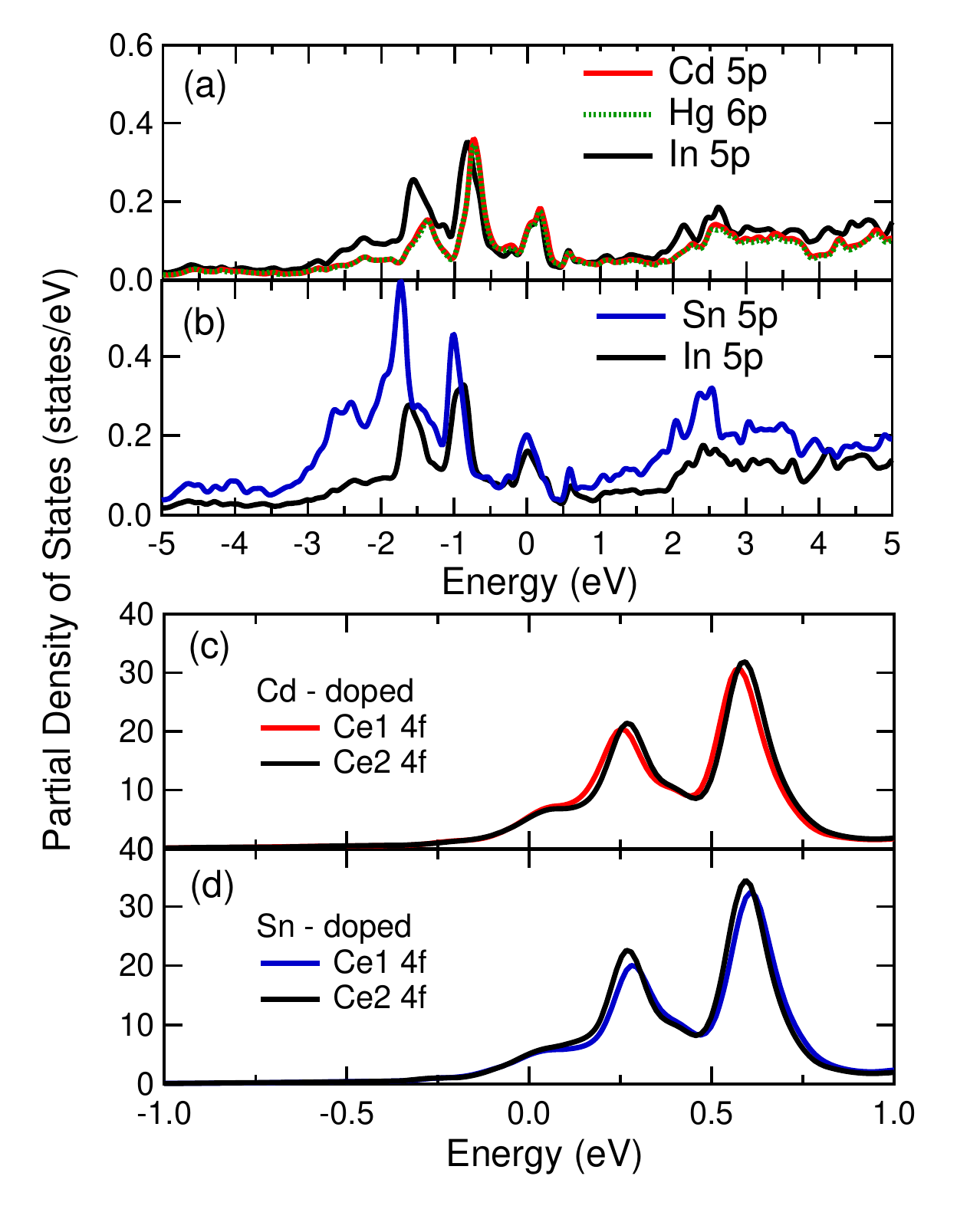}
\caption{(Color online) DFT results from three supercell calculations of Cd, Sn, and Hg doping in CeCoIn$_5$. (a) Comparison of the PDOS of the Cd impurity to a well removed In(1) site. Also shown is the PDOS from an identical calculation with Hg (dotted line) overlapping with the Cd curve. (b) Same as (a) but for the calculation done with Sn doping. (c) and (d) PDOS of Ce1 atoms which neighbor the dopant atom, and Ce2 which are furthest from the Cd and Sn dopant atom, respectively.}\label{fig.3}
\end{centering}
\end{figure}

\begin{figure}[t!]
\begin{centering}
\includegraphics[width=0.45\textwidth]{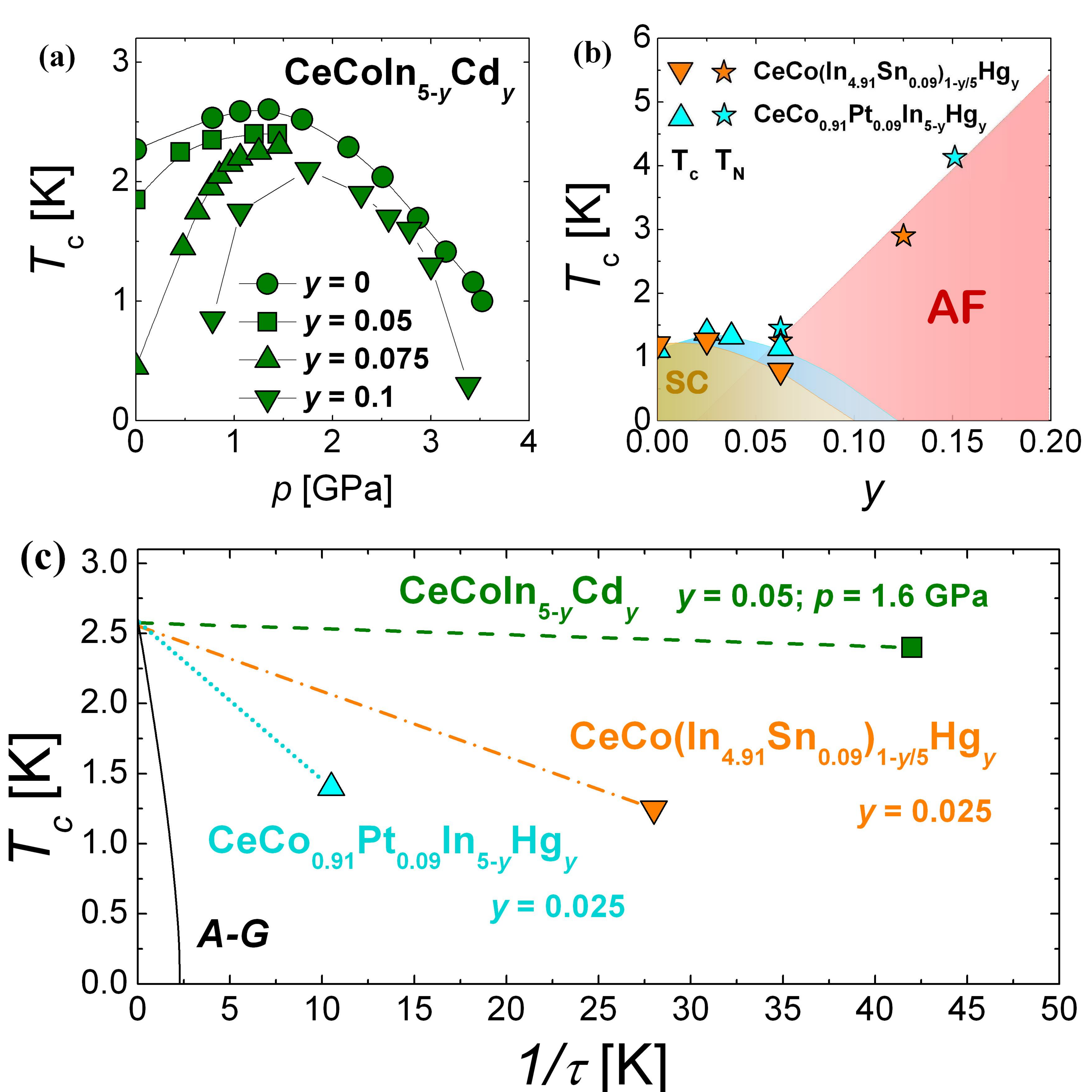}
\caption{(Color online) (a) Pressure dependence of $T_c$ of CeCoIn$_{5-y}$Cd$_{y}$ samples. (b) Temperature-composition phase diagram of CeCo$_{0.91}$Pt$_{0.09}$In$_{5-y}$Hg$_{y}$ and CeCo(In$_{4.91}$Sn$_{0.09}$)$_{1-\frac{y}{5}}$Hg$_{y}$. (c) $T_c$ suppression versus conservative estimates of the scattering rate for CeCoIn$_{4.95}$Cd$_{0.05}$, CeCoIn$_{4.88}$Sn$_{0.09}$Hg$_{0.026}$ and CeCo$_{0.91}$Pt$_{0.09}$In$_{4.97}$Hg$_{0.026}$ (see text). The latter two have had the contribution to the scattering rate and $T_c$ suppression from Hg subtracted to isolate the effects due to the electron dopants (Sn and Pt). The solid black line is the expectation for a $d$-wave SC based on AG theory.}\label{fig.4}
\end{centering}
\end{figure}

$T_{c}$ is affected simultaneously by pair breaking and electronic tuning. To isolate the pair breaking component, we utilize pressure and co-doping to eliminate the electronic tuning contribution. For hole doped samples with Cd it has been shown that pressure reversibly tunes the global magnetic phase diagram\cite{11}. Consequently, for any CeCoIn$_{5-y}$Cd$_{y}$ sample pressure can be used to tune to its optimal $T_{c}$ (see Fig.~\ref{fig.4}a). From this maximal $T_{c}$ with pressure the pair breaking component $\frac{dTc}{dCd}$~=~-5~K/Cd can be extracted \cite{22}.  We assume that the effect of Hg and Cd doping is the same. This is supported by almost identical phase diagrams and our DFT calculations in Fig.~\ref{fig.3}.

To isolate the pair breaking component for the electron doped samples with Sn and Pt we utilize co-doped samples with Hg for which we now know the pair breaking component. We have chosen Pt and Sn doped CeCoIn$_{5}$ samples at concentrations such that $T_{c}$~$\sim$~1.2~K ($x$~=~0.09). Then we have doped these two systems with Hg (substitution on the In site) which is a hole dopant. Fig.~\ref{fig.2}c demonstrates that the coherence temperature is reversibly tuned by this procedure. Moreover, for both systems upon Hg doping, superconductivity initially increases (see Fig.~\ref{fig.1}) - clear evidence of the reversible electronic tuning - and forms a superconducting dome seen in Fig.~\ref{fig.4}b. For higher Hg concentrations long-range antiferromagnetic ordering occurs as observed previously in CeCoIn$_{5-x}M_{x}$ studies where $M$~=~Hg or Cd\cite{11,12,13,14}. Note that the extra electrons from doping with Pt or Sn is nearly perfectly compensated by the holes from the Hg dopants with respect to the global phase diagram and $T^{*}$. The maximal $T_{c}$ in the co-doped samples with $x$~=~0.09 occurs at $y$~=~0.025. At this point electronic tuning is again optimized. The expected decrease in $T_{c}$ due to Hg doping is only 0.125~K. The remaining $T_{c}$ suppression relative to the maximal $T_{c}$ of CeCoIn$_{5}$ under pressure ($T_{c}$~$\approx$~2.6~K\cite{mn}) can then be attributed to pair breaking by Pt and Sn. We find $\frac{dTc}{dPt}$~=~-11.2~K/Pt and $\frac{dTc}{dSn}$~=~-13.3~K/Sn. The fact that the pair breaking strength of Pt substituted on the Co site is comparable to Sn substituted primarily on the in-plane In site indicates a similar impurity potential effect independent of the dopant location. This demonstrates that the 115 structure can \textit{not} be thought of as 'active' CeIn$_3$ layers separated by $T$In$_2$ 'buffer' layers. Interestingly, the pair breaking strength of electron doping matches that of rare-earth substitution $\frac{dTc}{dR}$~$\approx$~-10~K/$R$ \cite{Petrovic}, possibly reflecting an equivalence between physically removing Ce moments and quenching the Ce moments through increased local hybridization, although we note that the normal state scattering rates are different.

These pair breaking rates reflect the remarkable robustness of superconductivity to non-magnetic impurities in Ce$T$In$_{5}$. To quantify this we estimate the scattering rate caused by our various dopants. $1/\tau$ = $ne^{2}\Delta\rho/m^{*}$ = $\Delta\rho/\mu_{0}\lambda^{2}$. From the inset of Fig.~\ref{fig.1} we can see that $\Delta\rho$ = 12, 14 and 9 $\mu\Omega$~cm for optimally tuned Cd = 0.05, (Sn = 0.09, Hg = 0.025), and (Pt = 0.09, Hg = 0.025) samples, respectively. Values of the penetration depth $\lambda$ range from 190 to 550 nm \cite{Ormeno, Ozcan, Higemoto}. Thus, the largest reported penetration depth gives conservative estimates for the minimum scattering rate. According to Abrikosov-Gorkov (AG) theory for a $d$-wave superconductor with unitary scattering of non-magnetic impurities the initial rate of $T_c$ suppression should be $dT_{c}/d(1/\tau)$ = -$\pi$/4 $\approx$ -1\cite{AG}. Our conservative estimates for the maximal $dT_{c}/d(1/\tau)$ are -0.006, -0.04, and -0.09 for Cd, Sn, and Pt doping, respectively. To obtain the values for Sn and Pt doping we have subtracted Hg's contribution to the scattering rate in the co-doped samples in the same way as the $T_c$ suppression was determined above. As shown in Fig.~\ref{fig.4}c this demonstrates that \textit{the pair breaking by nominally non-magnetic dopants in CeCoIn$_{5}$ is more than one order of magnitude weaker than expected on the basis of AG theory for d-wave superconductors}.

Anomalously small $T_{c}$ suppression from nominally non-magnetic impurities in pnictides has controversially been argued to imply a conventional order parameter which does not change sign\cite{Kontani, HirschfeldRPP}. Meanwhile, weak $T_c$ suppression by non-magnetic impurities in cuprates has led to the realization that this can arise from small coherence lengths leading to spatial inhomogeneity, anisotropic scattering, strong coupling superconductivity, and induced magnetic moments \cite{Franz, Haran, Kulic, Monthoux}. Many of the explanations for cuprates are likely applicable here as well. CeCoIn$_5$ is a well established strong-coupling $d$-wave superconductor with a short coherence length\cite{2}. Spatial inhomogeneity by dopant atoms in CeCoIn$_5$  reveals the extended nature of the impurities\cite{12,7}. For instance, Cd dopants are known to create antiferromagnetic droplets\cite{12} similar to the case of Zn doping in cuprates \cite{all,JXZ}.

In summary, we have observed  the effects of both reversible electronic tuning and superconducting pair breaking in the unconventional superconductor CeCoIn$_{5}$ driven by electron (Pt and Sn) and hole (Hg, Cd) doping. DFT calculations reveal the doping dependence is qualitatively understood from the local increase (decrease) in hybridization caused by Sn (Cd) dopants. In addition, the equivalence of doping with Cd and Hg is demonstrated, thereby illustrating the utility of DFT calculations for understanding doping trends in heavy fermions. We have further shown that pair breaking by electron dopants is roughly independent of their location within the unit cell and is stronger than with hole dopants which induce antiferromagnetic droplets. Conservative estimates for the minimal scattering rate reveal that we can add CeCoIn$_5$ to the list of unconventional superconductors for which $T_c$ suppression by non-magnetic impurities is significantly weaker than expected by Abrikosov-Gorkov theory.

\begin{acknowledgments}

We gratefully acknowledge fruitful discussions with T.~Durakiewicz, M.~Graf and J.~Paglione. Work at Los Alamos National Laboratory was performed under the auspices of the U.S. Department of Energy, Office of Science. Z.F acknowledges support from NSF Grant No. DMR-0801253. T.P acknowledges support from the NRF grant funded by the Ministry of Education, Science and Technology (MEST) (No. 2010-002672).
\end{acknowledgments}

\end{document}




\title{Supplemental Material for Electronic Tuning and Uniform Superconductivity in CeCoIn$_{5}$}

\author{K. Gofryk$^{1}$}
\email{gofryk@lanl.gov}
\author{F. Ronning$^{1}$}
\email{fronning@lanl.gov}
\author{J.-X.~Zhu$^{1}$}
\author{M.~N.~Ou$^{1,2}$}
\author{P.~H.~Tobash$^{1}$}
\author{S.~S.~Stoyko$^{3}$}
\author{X.~Lu$^{1}$}
\author{A.~Mar$^{3}$}
\author{T.~Park$^{4}$}
\author{E.~D.~Bauer$^{1}$}
\author{J.~D.~Thompson$^{1}$}
\author{Z.~Fisk$^{5}$}
\affiliation{$^{1}$Los Alamos National Laboratory, Los Alamos, New Mexico 87545, USA\\$^{2}$Institute of Physics, Academia Sinica, Taipei, Taiwan\\$^{3}$Department of Chemistry, University of Alberta, Edmonton, Alberta T6G 2G2 Canada\\$^{4}$Department of Physics, Sungkyunkwan University, Suwon 440-746, South Korea\\$^{5}$Department of Physics and Astronomy, University of California, Irvine, California 92697, USA}

\date{\today}

\maketitle

\textbf{In this supplemental section we describe the details of our density functional theory calculations and also elaborate in greater detail the procedure we have applied to isolate pair breaking strength in the CeCoIn$_{5-y}$Cd$_{y}$ system.}\\

\begin{figure}[b]
\begin{centering}
\includegraphics[width=0.4\textwidth]{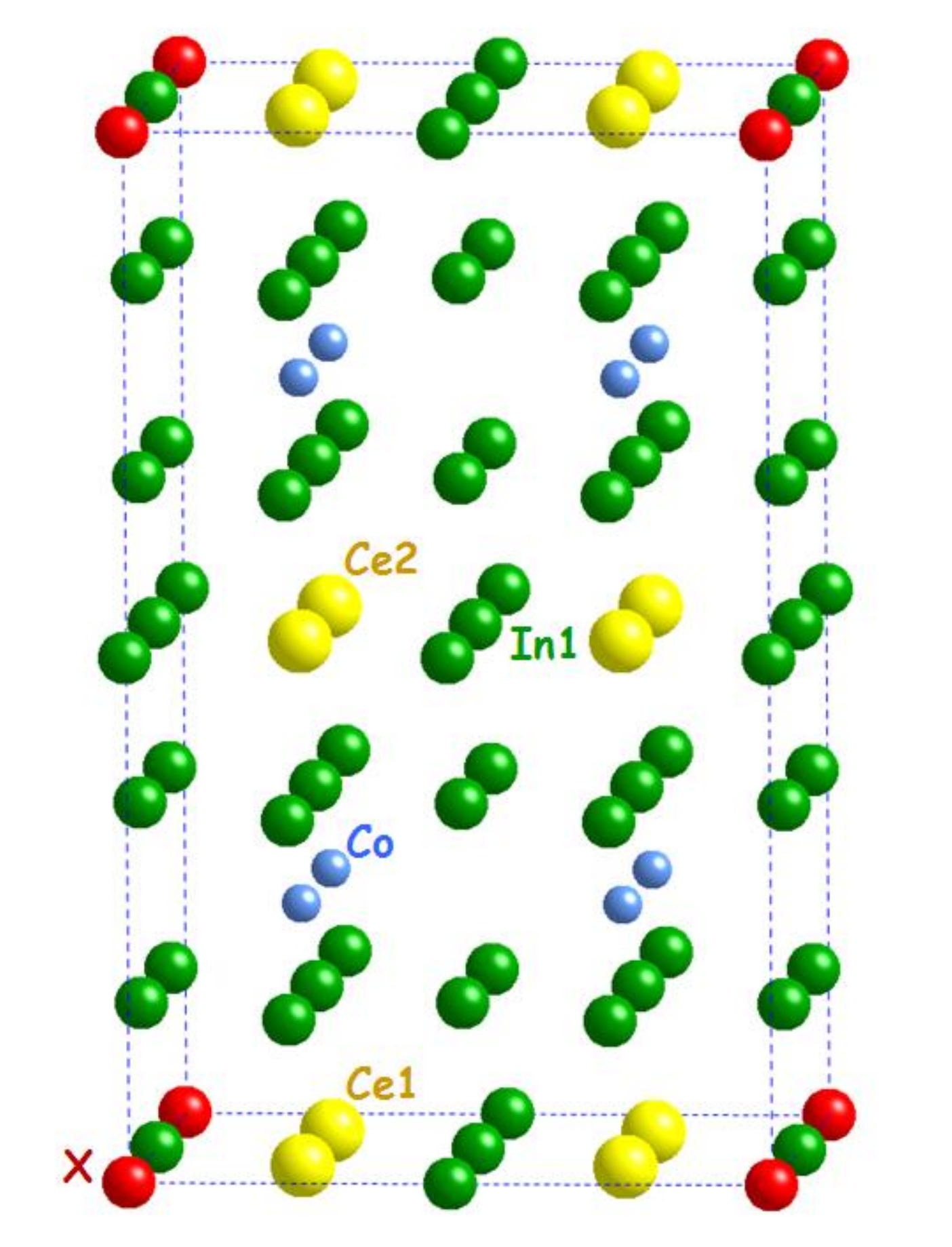}
\caption{(Color online) Structure of the 2x2x2 supercell used in the DFT calculations.}\label{figS1}
\end{centering}
\end{figure}

We use the WIEN2k code to calculate the electronic structure using the generalized gradient approximation with the Perdew-Burke-Ernzerhof exchange correlation potentials \cite{wien,PBE}. Spin-orbit coupling for the Ce $f$-states is included with a second variational method. We use a 2x2x2 supercell as shown in Fig.~\ref{figS1}. The experimental lattice parameters for pure CeCoIn$_{5}$ are used for all calculations as measurements reveal no changes for any of the dopants studied \cite{Sn,Hg}. We assume the dopant atoms sit exclusively on the In(1) sites as EXAFS measurements reveal that Cd, Sn, and Hg preferentially reside on the In(1) site \cite{Daniel}. In this way, we can also minimize the total doping as the In(2) site has a multiplicity of 4 compared with the multiplicity of 1 for the In(1) site.

\begin{figure}[b]
\begin{centering}
\includegraphics[width=0.45\textwidth]{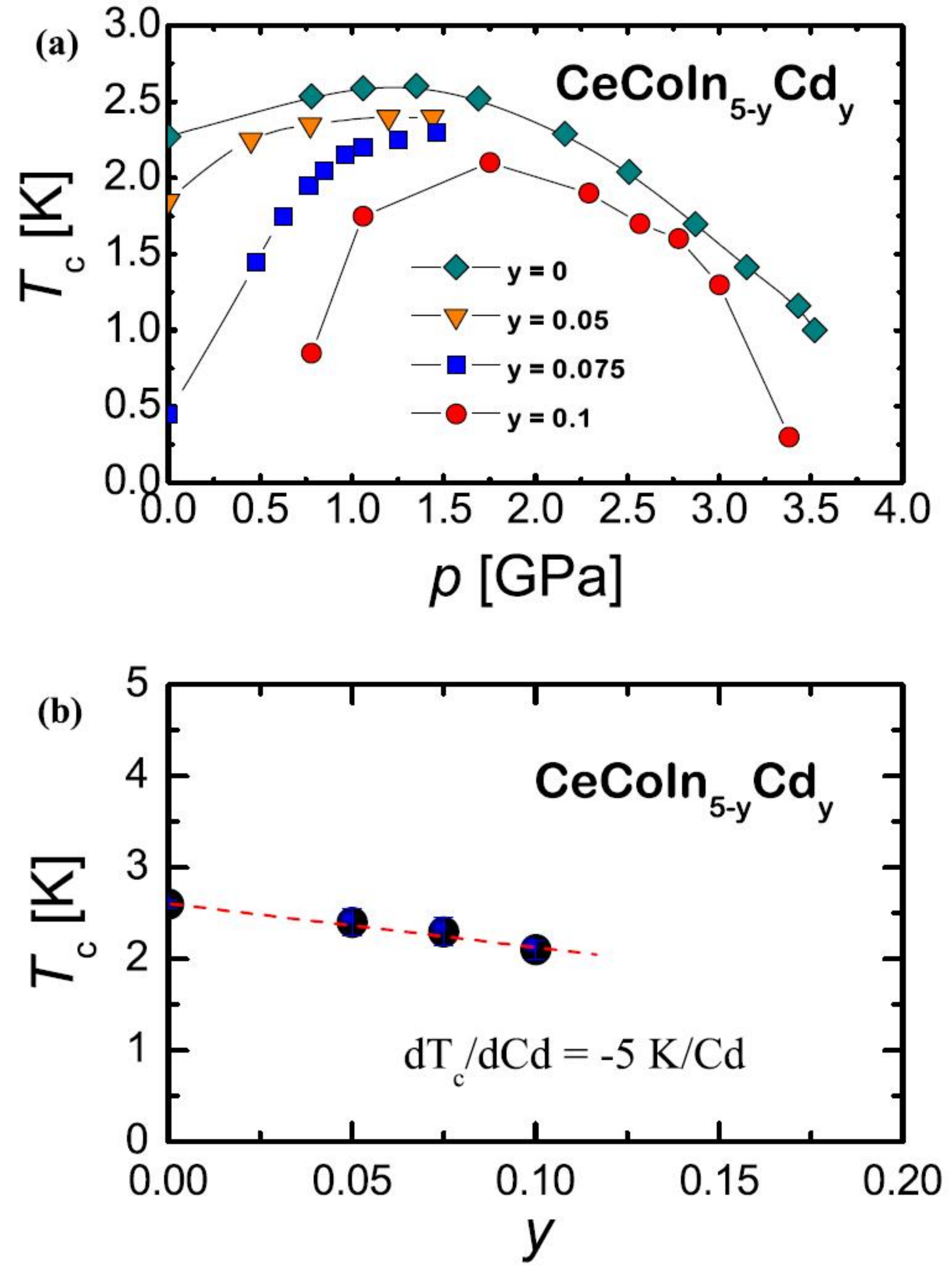}
\caption{(Color online) (a) Pressure dependence of $T_{c}$ of CeCoIn$_{5-y}$Cd$_{y}$ samples. (b) maximal $T_{c}$ vs. $y$ for CeCoIn$_{5-y}$Cd$_{y}$.}\label{figS2}
\end{centering}
\end{figure}

For these studies, single crystals of CeCoIn$_{5-y}$Cd$_{y}$ were grown using a standard In-flux method \cite{2,21} in which various amounts of Cd were added to the flux. The doping levels were determined by single crystal x-ray diffraction and microprobe analysis. For the samples the actual concentrations used here have been established to be 10~\% of the nominal concentrations \cite{11,13}. Pressure-dependent electrical resistance studies were carried out in a hybrid Be-Cu/NiCrAl clamp-type pressure cell that contains a Teflon cup filled with silicone as the pressure-transmitting medium, samples and a small piece of Sn, whose inductively measured $T_{c}$ served as a manometer. The standard 4-wire method has been used.

Cadmium doping influences the $T_{c}$ of CeCoIn$_{5}$ through both an electronic tuning effect and a pair breaking effect. We use pressure to eliminate the electronic tuning effect and isolate the pair breaking effect. Fig.~\ref{figS2}a (and Fig.~4a of the main text) shows the pressure dependence of the superconducting temperature for CeCoIn$_{5-y}$Cd$_{y}$ for different Cd concentrations. As may be seen $T_{c}$ initially increases under pressure and forms a broad superconducting dome. For each sample, the pressure where $T_{c}$ is maximal is the point at which the electronic tuning has been optimized. Since pressure does not add a pair breaking component the suppression of the optimized $T_{c}$ is attributed to the pair breaking strength of the Cd dopants. Taking the maximal $T_{c}$ for each doping we construct the optimal $T_{c}$ vs. $y$ plot as shown in Fig.~1b. As seen the optimized $T_{c}$ depression under Cd doping is -5~K/Cd (see dashed red line in Fig.~\ref{figS2}b).

For the case of Pt- and Sn-doped CeCoIn$_{5}$ samples we cannot use pressure to optimize $T_{c}$ since a negative pressure would be needed. Therefore we achieve the necessary electronic tuning in these materials by using co-doping by Hg. As seen from Fig.~4b in the paper the maximal $T_{c}$ for the co-doped samples with $x$~=~0.09 occurs at $y$~=~0.025. At this point electronic tuning is again optimized. Taking into account the so obtained scattering strength the expected decrease in $T_{c}$ due to Hg doping is expected to be (5~K/Cd $\times$ 0.025 Cd =) 0.125~K only. The remaining $T_{c}$ suppression relative to the maximal $T_{c}$ of CeCoIn$_{5}$ under pressure ($T_{c}$~$\approx$~2.6~K\cite{mn}) can then be attributed to pair breaking by Pt and Sn. We find $\frac{dTc}{dPt}$~=~-11.2~K/Pt and $\frac{dTc}{dSn}$~=~-13.3~K/Sn.